\documentclass[10pt]{article}

\usepackage{amsmath}
\usepackage{amssymb}
\pdfoutput=1
\usepackage{graphicx}
\usepackage{graphicx,epstopdf}
\usepackage{subfigure}
\usepackage{subfig}
\usepackage{float}
\usepackage{cite}

\usepackage{color}
\usepackage{soul}
\usepackage{upgreek}

\usepackage{longtable}

\usepackage{setspace}
\usepackage{booktabs}

\usepackage{times}
\usepackage{mathptmx}

\usepackage{indentfirst}
\usepackage{bm}

\usepackage{url}

\usepackage[labelfont=bf,labelsep=period,justification=raggedright]{caption}

\usepackage[numbers,sort&compress]{natbib}
\makeatletter
\renewcommand{\@biblabel}[1]{\quad#1.}

\makeatother

\date{}

\begin{document}		
	\begin{flushleft}
		{\Large \textbf {Quantifying the attractor landscape and transition path of distributed working memory from large-scale brain network}}
		\\
		\textbf{Authors:} Leijun Ye$^{1,2}$, Chunhe Li$^{1,2,3,\ast}$
		\\
		\textbf{Affiliations:} 1 Institute of Science and Technology for Brain-Inspired Intelligence, Fudan University, Shanghai, China
		\\
		2 Shanghai Center for Mathematical Sciences, Fudan University, Shanghai, China
		\\
		3 School of Mathematical Sciences, Fudan University, Shanghai, China
		\\
		$^*$Corresponding author
		\\
		\textbf{E-Mail:} chunheli@fudan.edu.cn
	\end{flushleft}
	
	\section*{Abstract}
	
	Many cognitive processes, including working memory, recruit multiple distributed interacting brain regions to encode information, rather than being associated with the function of a single brain region. How to understand the underlying cognition function mechanism of working memory is a challenging problem, which involves neural circuit configuration from multiple brain regions as well as stochastic transition dynamics between brain states. The energy landscape idea provides a tool to study the global stability and stochastic transition dynamics in the distributed cognitive function system. However, how to quantify the energy landscape in a realistic large-scale brain network remains unclear. Here, based on an anatomically constrained computational model of large-scale macaque cortex, we quantified the underlying multistable attractor landscape of distributed working memory. In the absence of external stimulation, the landscape exhibits three stable attractors, a spontaneous state, and two memory states. In the attractor landscape framework, the working memory function is governed by the change of landscape topography and the switch of system state according to the task requirement. The barrier height inferred from landscape topography quantifies the global stability of memory state and robustness to non-selective random fluctuations and distractor stimuli. The kinetic transition path identified by the minimum action path approach reveals that the spontaneous state serves as an intermediate state during the switch between the two memory states, the memory stored in the cortical area with higher hierarchy is more stable, and information flow follows the direction of hierarchical structure. These results provide new insights into the underlying mechanism of distributed working memory function, and the landscape and kinetic path approach can be applied to other cognitive function-related problems in brain networks.
	
	\textbf{Key words}  Distributed working memory, Mathematical model, Stochastic neurodynamics, Energy landscape, Transition path

	\section{Introduction}
	The brain is a complex system performing multiple cognitive functions~\cite{Varela2001NRN}. Working memory refers to the brain's ability to retain and manipulate information without external inputs in a short time period, which is a basic ingredient to many other cognitive functions, such as learning and decision-making. Traditionally, working memory is related to the persistent activity during a mnemonic delay in localized brain regions, such as the prefrontal cortex (PFC)~\cite{Cohen1997Nature,Inagaki2019Nature,Riley2016frontiers}. Recently, many studies revealed that persistent neural activity during memory delays can be found in a range of specialized brain areas, from sensory to parietal and prefrontal cortex~\cite{Christophel2017TCS}, suggesting the distributed nature of working memory~\cite{Roy2022NC}. However, how these distributed brain regions interact with each other and achieve cognitive function cooperatively remains to be elucidated.
	
	The brain faces a vast repertoire of ever-changing external and internal stimuli. To respond to these stimuli with appropriate behaviors, the brain dynamics should switch between several possible network states dynamically, rather than settle in one single strong states~\cite{Golos2016PlosCB}. Some studies identified the rapid transitions between a few discrete functional connectivity states of brain activity using the resting-state functional Magnetic Resonance Imaging (fMRI)~\cite{Hansen2015NeuroImage,Cabral2017NeuroImage}. The electro- or magnetoencephalography recordings also provide evidence for the multistability in brain rhythms~\cite{Freyer2009JN,Freyer2012PlosCB}. Similar transition dynamics have also been observed in the kinetic model of gene regulation~\cite{Ye2021JCP,Li2019JCP}, suggesting the universality of multistability across a broad class of biological phenomena. Then, a fundamental problem arises as to how the transition between different brain states occurs, which is one of the most essential questions in system neuroscience~\cite{Kringelbach2020CP}. Due to the complexity of brain states, the in-depth study of state transitions has been minimally successful.
	
	Drawing on insights from dynamical system theory, brain states can be considered as attractors or the stable states of interacting brain regions. For the working memory function, memory can be considered as being stored in the corresponding attractor~\cite{Wimmer2014NN,Seeholzer2019PlosCB,Wang2021TIN,Inagaki2019Nature}. Experiments have provided direct evidence for multiple discrete attractors as the underlying mechanism of different motor directions decision in a delayed response task~\cite{Inagaki2019Nature}. Besides, attractor dynamics also apply in decision-making and attention~\cite{Wong2006JNeurosci,Deco2005PIN,Zhang2019bioRxiv}. The stability of attractors is also related to some disorders of cortical function such as depression~\cite{Rolls2016NBR} and schizophrenia~\cite{Rolls2021TP}.
	
	Recently, an anatomically constrained computational model has been proposed by Mejias \textit{et al.} to elucidate the mechanism of distributed working memory in a large-scale network of macaque neocortex~\cite{Mejias2022Elife}. In their study, the brain regions showing self-sustained persistent activity during the delay period agree well with the meta-analyses results~\cite{Leavitt2017}. However, the mechanism of distributed working memory was explored by tracing the neural activity trajectories in a finite timescale, which mostly reflects the local property rather than the global characteristics of the dynamical working memory system. So is there any possible way to obtain the global natures of collective behaviors emerging in complex neural systems? Inspired by thermodynamics and statistical physics, the collective properties of a complex working memory system composed of multiple interacting elements can be explored by the thermodynamics quantities, such as free energy and entropy. In neuroscience, the concept of "energy function" was firstly proposed by Hopfield to explore the computational properties of neural circuits, providing a clear dynamical picture of how memory storage and retrieval are implemented~\cite{Hopfield1982,Hopfield1986}. Energy landscape was also used in whole-brain modeling to examine the structure-function relationship in the brain~\cite{Zhang2022NeuroImage}. In systems biology, the Waddington landscape has been used as a metaphor for cell development and differentiation~\cite{li2013PLOS,Shakiba2022CellSystems}. The energy landscape provides a quantitative and intuitive way to investigate the global and transition properties of complex systems.
	
	Some efforts have been devoted to quantifying the energy landscape of the working memory system with a realistic neural network, highlighting the importance of landscape topography in determining the function of the neural network and uncovering the essential stability-flexibility-energy
	trade-off in working memory~\cite{Yan2013PNAS,Yan2020PlosCB}. Besides, for decision-making function, the energy landscape advances our understanding of the dynamical mechanism of change of minds~\cite{Ye2021Frontiers}. However, these attempts are based on the local neural circuits. For the high-dimensional distributed working memory system with multiple interacting brain areas, how to quantify the landscape and visualize it remains challenging. Furthermore, our brain is inherently noisy. For the storage of working memory, the self-sustained persistent activity during a mnemonic delay is desired to be robust to noise. However, the general principles for maintaining robust working memories in noisy biological systems are unclear.
	
	To address these questions, in this work, based on the anatomically constrained computational model of distributed working memory described by 90-D ordinary differential equations (ODEs)~\cite{Mejias2022Elife}, we quantified the multistable attractor landscape with our developed landscape quantification and visualized it with dimension reduction. We explored the state transition in working memory function by barrier height inferred from landscape topography and kinetic transition path by minimum action path (MAP) approach. In the regime where none of the isolated areas is capable of generating self-sustained activity and when no stimulus is applied, the whole neocortex exhibits three different attractors, i.e., the resting state with spontaneous low activity and two stimulus-selective memory states. The transition path between the two memory states obtained from MAP not only goes through the spontaneous state but also shows sequential deactivation from sensory areas to association areas, which is consistent with the cortical hierarchy defined by anatomical studies. In addition, we found that the barrier height separating the two memory attractor basins increases for non-selective ramping input to the network, explaining experimental observations showing that the robustness to distractors is stronger during the late-delay epoch compared to the early-delay epoch in the working memory task. Furthermore, in the regime where some of the isolated areas operate in a multistable regime, more attractors emerge, potentially subserving internal processes. We also explored the role of the network structure on the robustness against random fluctuations.
	
	
	\section{Results}
	\subsection{Energy landscape construction from large-scale working memory model}
\begin{figure}[htbp]
		\centering
		\includegraphics[width=\linewidth]{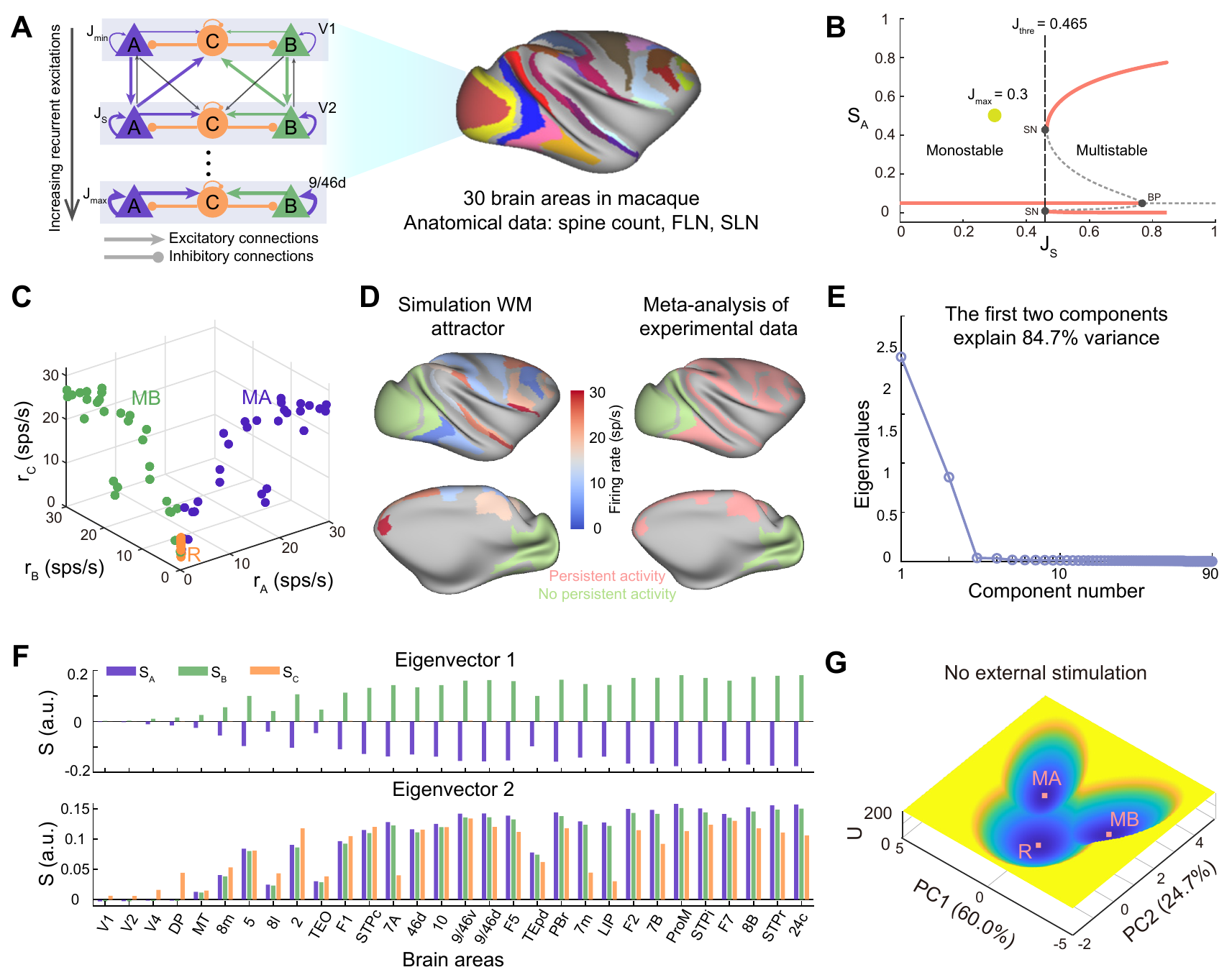}
		\captionsetup{font={stretch=1}, justification=raggedright}
		\caption{\label{fig:Fig.1} (A) The schematic diagram of the anatomically constrained computational model of distributed working memory in macaque with 30 brain areas. Each brain area is modeled by two selective excitatory neural populations, labeled as A and B, and one inhibitory population, labeled as C. Each excitatory population has self-excitations and inhibits each other mediated by population C. To introduce the area-to-area heterogeneity, the recurrent strength of selective populations and coupling strength from the excitatory population to the inhibitory one are calibrated by the gradient of dendritic spine count. The inter-areal projections are based on quantitative connectomics data, the fraction of labeled neurons (FLN), and supragranular layered neurons (SLN). For clarity, the cross-couplings between two excitatory populations are not shown. (B) The bifurcation diagram with respect to the recurrent strength ($J_{S}$) for the isolated area. In this study, unless specified otherwise, we set the maximum recurrent strength ($J_{max}$) as 0.3, lower than the bifurcation point ($J_{S}=0.465$), so that all the brain areas are monostable when isolated. SN: saddle-node, BP: branch point. (C) The three fixed points of the distributed working memory model, resting state (R), memory state encoded by population A (MA), and memory state encoded by population B (MB). (D) There is a strong overlap between the persistent activity of the memory state predicted by the model and the meta-analysis of experimental data~\cite{Leavitt2017}. (E) The eigenvalues of the covariance matrix of the probabilistic distribution of system states for dimension reduction of the landscape. The first two components explain 84.7\% of the variance. (F) The eigenvectors correspond to the first two components. (G) The quantified tristable potential landscape after dimension reduction when no external stimulus is applied. Note that the landscape is symmetric along the PC1 axis. The parameters are chosen as default values.}
	\end{figure}

	The large-scale computational model of distributed working memory (DWM) function we employed here is firstly developed by ~\cite{Mejias2022Elife} (schematized in Fig.~\ref{fig:Fig.1}(A)). Different from previous modeling efforts which are restricted to the behavior of local circuits~\cite{Edin2009PNAS,Murray2017JN}, the DWM model is characterized by the multi-regional interactions supporting the emergence of distributed persistent activity in working memory function. Furthermore, the DWM model is constrained by the anatomical data, including the dendritic spine count values of layer 2/3 pyramidal neurons across different cortical areas~\cite{Elston2007}, the fraction of labeled neurons (FLN) and the supragranular layered neurons (SLN)~\cite{Markov2013Science}. Each cortical area is modeled as two selective excitatory neural populations, labeled as A and B, and one inhibitory neural population, labeled as C~\cite{Wong2006JNeurosci}. This subnetwork is fully connected and the winner-take-all competition between populations A and B mediated by GABAergic inhibitory population C results in the stimulus-selective self-sustained persistent activity~\cite{Wong2006JNeurosci}. The local circuit is placed in 30 areas across all four neocortical lobes and all the cortical areas are interconnected according to the neuroanatomical connectivity matrix FLN. The cortical hierarchy is estimated from the laminar connectivity data, SLN~\cite{Markov2014JCN}. A macroscopic gradient of synaptic excitations is introduced into the model~\cite{Wang2020NRN}, that is, both the local recurrent and long-range excitations are area-specific and increase along the cortical hierarchy. In particular, the maximum strength of local and long-range excitations for the area at the top of the cortical hierarchy is denoted by $J_{max}$. Mathematically, the DWM model can be described by 90-dimensional ordinary differential equations (ODEs): $\frac{d \bm{x}(t)}{dt} = \bm{F}(\bm{x}(t))$ with $\bm{x}$ represents the system variables and $\bm{F}(\bm{x})$ denotes the driving force (see Methods for details). These ODEs describe the temporal evolution of synaptic current of 90 populations (30 brain areas with 3 populations in each area).
	
	Firstly, we explore the dynamics of each local circuit, which is also the classical working memory model. From the bifurcation diagram with respect to recurrent excitations ($J_{S}$) in Fig.~\ref{fig:Fig.1}(B), we found that there exists a threshold level ($J_{S}=0.465$) separating the monostable dynamics from the multistable dynamics. The monostability means that the system has one exclusively resting state with all the population activity at a low level. Multistability refers to three coexisting stable states, one resting state, and two stimulus-selective persistent activity states with one of the selective populations at high firing activity. In this work, unless specified otherwise, we set $J_{max}=0.3$ lower than the threshold level, that is, all the cortical areas are monostable when isolated. In this situation, the emergence of persistent activity pattern is due to the interactions between cortical areas.
	
	Then we consider the dynamics of distributed working memory function engaging multiple interacting brain regions. The brain is inherently noisy, originated from intrinsic random fluctuations and external distraction~\cite{Deco2009PIN}. Previous work analyzed individual dynamical trajectories in the finite time scale as well as the phase transition using the bifurcation approach~\cite{Mejias2022Elife}. However, the stochastic transition properties and global stability of working memory systems remain to be quantified. Here, we focus on the probabilistic evolution of the working memory system to explore corresponding stochastic dynamics. Since neural network systems are open systems constantly interacting with the external environment, we resort to the non-equilibrium statistics physics approach. With our developed landscape and transition path framework (see Methods for details)~\cite{Li2014PNAS,li2013PLOS,Kang2021AS}, we can quantify the potential landscape of the attractor dynamics of working memory function and further study the probabilistic switching process across barriers on the potential landscape to describe the transition dynamics.
	
	Starting from the DWM model, the Langevin equation describing the stochastic dynamics of working memory system takes the forms as $\frac{d \bm{x}(t)}{dt} = \bm{F}(\bm{x}(t)) + \bm{\zeta}(t)$ with $\bm{\zeta}$ assumed to be Gaussian white noise. The Langevin equation is corresponding to a Fokker-Plank diffusion equation describing the temporal evolution of system probability distribution. It is hard to solve the diffusion equation directly, due to the high dimensionality and nonlinear interactions of the DMW model. By assuming the probabilistic distribution of the system state is Gaussian and determining the mean and variance of Gaussian distribution from moment equations, the temporal evolution of the probabilistic distribution of the system state can be solved numerically (see Methods and Supplementary Fig.~1 for details). Then, the potential landscape is defined as $U(\mathbf{x})=-\ln P_{s s}(\mathbf{x})$~\cite{Li2014PNAS} with $P_{ss}$ representing the probabilistic distribution at steady state. Furthermore, for the visualization of the high-dimensional landscape, we conducted dimension reduction to project the landscape onto the subspace spanning by the first two components~\cite{Kang2021AS}.
	
	By initializing the DWM model with 10000 different initial conditions, the DWM system shows three stable states (Fig.~\ref{fig:Fig.1}(C)). Here, three different colors indicate three individual stable states. For each stable state, each dot represents one brain area, and its position in three-dimensional space is determined by the stable firing rate of corresponding three populations ($r_A$, $r_B$, and $r_C$). The orange state is identified as the resting state (R) since all three populations in each area are at the low spontaneous activity. The purple and green states are symmetrical and identified as the memory state encoded by population A (MA) and the memory state encoded by population B (MB), respectively. For MA, population A and population C of some specific areas show high firing rate while population B keeps in low activity. MB is on the contrary. Of note, the areas displaying activation in MA or MB involve frontal, temporal, and parietal lobes, except early visual areas (V1, V2, V4, DP, MT). This is consistent with the qualitative meta-analysis results concluding that persistent activity during the delay period in working memory tasks is more frequently observed in association areas while reports on early sensory are rare~\cite{Leavitt2017} (Fig.~\ref{fig:Fig.1}(D)).
	
	To study the global stability of multiple attractors, we mapped out the energy landscape of the working memory system and visualized the landscape in two-dimensional space from dimension reduction (Fig.~\ref{fig:Fig.1}(G))~\cite{Kang2021AS}. For the dimension reduction of the landscape, the first two principal components ($PC1$ and $PC2$) have a total contribution rate of 84.7\% (Fig.~\ref{fig:Fig.1}(E)), suggesting that projection on PC1 and PC2 preserves well the stability information of the original system. Fig.~\ref{fig:Fig.1}(F) presents the eigenvectors corresponding to the first two eigenvalues, respectively. For eigenvector 1, all $S_A$ in 30 brain areas are less than zero, $S_B$ are larger than zeros and $S_C$ are nearly zero. So $PC1$ reflects which selective population (A or B) shows high firing rate. Eigenvector 2 mainly indicates whether the inhibitory populations C are at a high firing level or not. The potential landscape of distributed working memory system after dimension reduction exhibits three attractors in the absence of external input (Fig.~\ref{fig:Fig.1}(G)). Each attractor is surrounded by its own basins of attraction (the dark blue region) and basins are separated by basin boundaries. By combining the characteristics of two eigenvectors and high-dimensional stable states, it is easy to infer which high-dimensional stable state the three low-dimensional attractors after dimension reduction correspond to, respectively. The basin with both $PC1$ and $PC2$ approximating to zero is identified as the R state. The remaining two symmetrical attractors are memory states. Basin with $PC1 < 0$ is for MA and $PC1 > 0$ for MB. Compared to the resting state, populations C in 30 brain areas show high firing rate in both two memory states, suggesting the winner-take-all competition between two excitatory populations is mediated by the inhibitory population. Of note, even without external input, the presence of noise can drive the probabilistic jumping across barriers between attractors, leading to the state transition between different attractors. Thus, the landscape provides a framework to explore stability and the stochastic transition dynamics of working memory function.
	
	\subsection{Landscape results explain the working memory function with multistable attractors}
	To meet the demands of external cognitive tasks, the working memory system needs to reconfigure itself, and the system state changes over time\cite{Taghia2018NC}. For example, a new stimulus can destabilize the present attractor and may drive the system to another attractor corresponding to the incoming stimulus. Furthermore, the failure to engage the optimal system state in a timely manner is associated with poorer task performance\cite{Taghia2018NC,Inagaki2019Nature}. In the attractor landscape framework, we can explicitly quantify the underlying potential landscape performing the distributed working memory function in a global way.
	
\begin{figure}[htbp]
		\centering
		\includegraphics[width=1\linewidth]{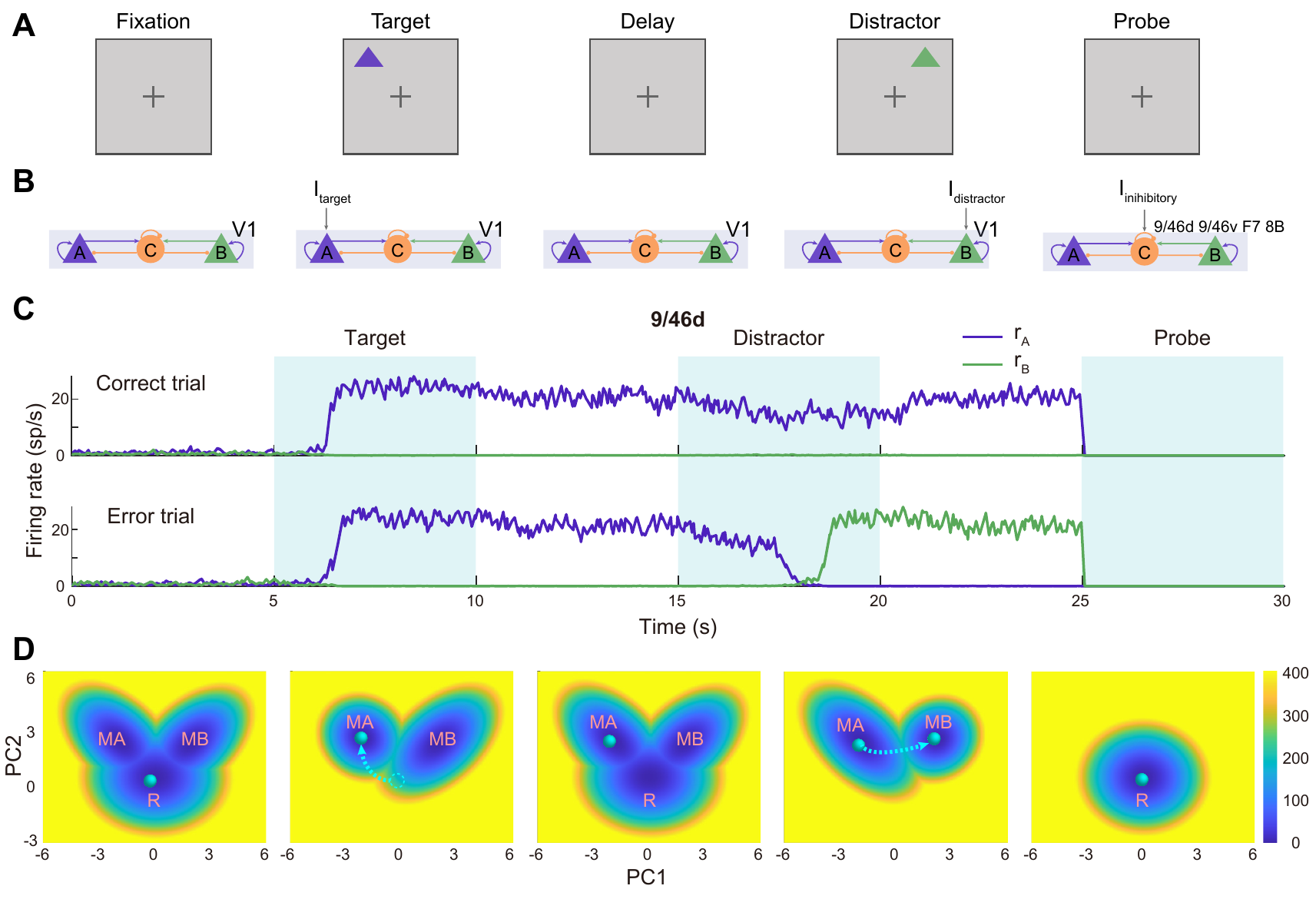}
		\captionsetup{font={stretch=1}, justification=raggedright} 
		\caption{\label{fig:Fig.2} (A-B) The schematic illustration of the visual working memory task. During the fixation phase, no external stimulus is applied. During the target phase, a cue is given, which is modeled as the population A of V1 receiving external current. Then the stimulus is removed and the target is expected to be maintained in working memory against random fluctuations and distractors encoded by population B at the delay epoch. Once the task has been done, the sustained activity should be shut down by delivering excitatory input to the inhibitory populations in four areas (9/46d, 9/46v, F7, 8B), leading to the clearance of working memory. (C) The stochastic neural activity traces of selected areas for correct and error trials (See Supplementary Material for behaviors of all areas). For correct trail, the target stimulus is maintained at delay epoch against distractor. On the contrary, the distractor induces the transition from stimulus-selective high activity state to the distractor-selective high state for error trail. (D) The corresponding attractor landscapes during different phases of the working memory task. The green ball represents the system state. Before the stimulus onset, the system stays at the resting state (R) with all the populations at low activity. The stimulus changes the landscape topography from tristability to bistability. And the system state transits from R to the dominated target-related memory state (MA). Even after the withdrawal of stimulus and the landscape topography returns back to tristability, the system state keeps staying at MA. However, the presentation of distractor stimulus changes the landscape again and may or may not induce the transition to distractor-related attractor (MB), corresponding error and correct trials, respectively.}
	\end{figure}

	We modeled the different phases of the working memory task (Fig.~\ref{fig:Fig.2}(A)) with the DWM model (Fig.~\ref{fig:Fig.2}(B)). During the target phase, a visual cue is given, which is modeled as the population A of V1 receiving external current ($I_{stimulus}$)~\cite{Mejias2022Elife}. Then the stimulus is removed and the target is expected to be maintained in working memory against random fluctuations and distractors encoded by population B ($I_{distractor}$) at the delay epoch. Once the action has been performed, the sustained activity should be shut down by delivering input ($I_{inhibitory}$) to the inhibitory populations in the top four areas of the hierarchy (9/46d, 9/46v, F7, 8B)~\cite{Mejias2022Elife}, leading to the clearance of working memory. In Fig.~\ref{fig:Fig.2}(C-D), dependent on the phase of the task, the dynamics of distributed working memory system are described by the time-evolving firing rate of neural populations in representative brain areas 9/46d (see Supplementary Material for trajectories in more areas) and the corresponding potential landscape. And the green ball represents the state of the working memory system.
	
	We argue that the landscape provides a quantitative and global explanation for the mechanism of working memory function at different stages through state transition dynamics (Fig.~\ref{fig:Fig.2}D). In the fixation period, since no stimulus is applied, the system stays at the resting state with all the populations at the low activity (Fig.~\ref{fig:Fig.2}(D), the first column). In the target period, the target visual stimulus changes the landscape topography from tristability to bistability and the symmetry of the landscape is broken. The MA state becomes dominated by deeper basins of attraction while the basin of the R state disappears (Fig.~\ref{fig:Fig.2}(D), the second column). When the stimulus is strong enough, the system transits from R to the MA state. Even after the withdrawal of stimulus in the delay epoch and the landscape returns back to tristability, the system can be maintained at the MA state since the MA is a stable attractor (the third column of Fig.~\ref{fig:Fig.2}(A-B, D)). So, the landscape attractor guarantees the stability of the working memory state. However, the distractor stimulus can change the landscape topography again and the MB state becomes dominated ((Fig.~\ref{fig:Fig.2}(D), the fourth column). Then there will be two outcomes. The system can be either distracted, switching to MB, or not distracted, keeping staying at MA, corresponding to error and correct trials, respectively ((Fig.~\ref{fig:Fig.2}(C)). At the end of the task, the inhibitory input changes the landscape topography to monostability and the system state returns back to the resting state (last column of Fig.~\ref{fig:Fig.2}(A-D)).
	
	\subsection{Transition paths characterize the dynamical switching process in working memory}
	For the multistable systems including the distributed working memory network, the system may hop between different stable states driven by external input and noise, which corresponds to the alteration of cognitive states~\cite{Deco2019PNAS,Taghia2018NC,Kringelbach2020CP}. For the working memory task, the switch from R to MA signifies the formation of target-related memory and the switch from MA to MB represents the alteration of memory to the distractor-related state.
	
\begin{figure}[htbp]
		\centering
		\includegraphics[width=\linewidth]{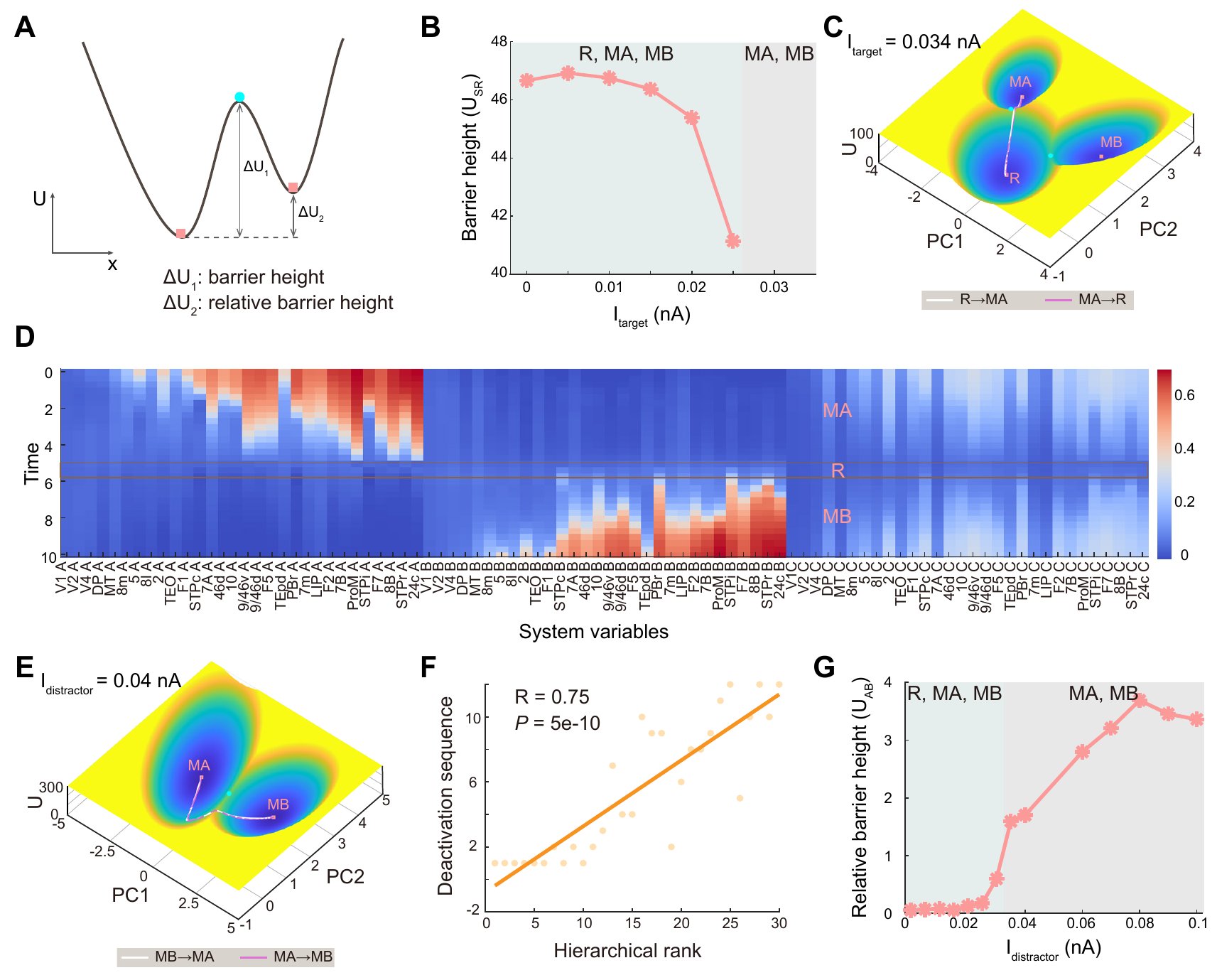}
		\captionsetup{font={stretch=1}, justification=raggedright}
		\newpage
		\caption{\label{fig:Fig.3} (A) The illustration for the definition of barrier height (potential difference between the potential at local minimum and saddle point) and relative barrier height (the difference between the two barrier heights) to quantify the stability of attractors. If the system tries to escape from the current attractor, it needs to cross the corresponding barrier. (B) To quantify the difficulty of state transition for the formation of memory, we define the barrier height $U_{SR}$ as the potential difference between the $U_{saddle}$ (the potential at the saddle point between R and MA state) and the $U_{min}$ (the potential at the minimum of R state) for $I_{target}=0.034nA$. The $U_{SR}$ decreases for higher strength of target stimulus, suggesting the easier formation of memory. (C) The projected transition path with dimension reduction on the landscape between R state and MA state. The transition from R to MA represents the formation of memory and the reverse path represents the clearance of memory. (D) The high-dimensional transition path from MA to MB before dimension reduction for a given time $T=10$. The X-axis represents the 90 system variables and the Y-axis represents the time points along the transition path. (E) The transition paths between MA and MB for $I_{distractor}=0.04nA$, which pass through the intermediate state R. The transition from MA to MB signifies the change of memory. (F) The deactivation sequence of population A in 30 brain areas shows a high correlation with the anatomical hierarchy as defined by layer-dependent connections~\cite{Markov2014JCN}. This suggests that the information of distractors flows along the hierarchy, from early sensory areas to association areas. (G) The robustness against distractors during the delay epoch. The increase of relative barrier height between MA and MB as a function of the strength of distractors suggests that MA is becoming more unstable while MB is becoming more stable, thus decreasing robustness of the system to distractors. R: resting state, MA: memory state encoded by population A, MB: memory state encoded by population B.
		}
	\end{figure}

	The state transition is associated with the stability of the corresponding state. An advantage of the landscape approach is that the stability of each attractor can be quantified through the height of the separating barriers. We provide an illustration diagram for the definition of barrier height to quantify the stability of attractors (Fig.~\ref{fig:Fig.3}(A)). More specifically, the barrier height is defined as the difference between $U_{saddle}$ (the potential at the unstable saddle point separating the basins of attraction for two neighboring attractors) and $U_{min}$ (the potential at a local minimum of the basin of attraction), which reflects the depth of the basin or attractor. Further, we can define relative barrier height as the potential difference between two barrier heights to measure the relative stability between attractors.
	
	During the target epoch in the working memory task, we found that with the increase in the strength of the target stimulus, the landscape shows a phase transition from tristability (coexisting R, MA, and MB states) to bistability (coexisting MA and MB state). Barrier height $U_{SR}$ is defined as $U_{SR}=U_{saddle}-U_{R}$ ($U_{saddle}$ is the potential at the saddle point between R and MA states and $U_{R}$ represents the potential minimum at R state), quantifying the stability of R state. The decaying trend of $U_{SR}$ indicates that the R state becomes more and more unstable with the increase of target strength until disappears and becomes the unstable saddle point between MA and MB (Fig.~\ref{fig:Fig.3}(B)).
	
	To quantify the dynamical process for the working memory function, a natural question arises as to what is the biological path during state transition, which will include the information for activation order of different brain regions during the transition process. Mathematically, the most probable transition paths between stable states can be identified by minimizing the action functional associated with all possible paths connecting initial and final states over a specified time interval of transition~\cite{E2004,Zhou2008JCP,Heymann2008}. For the equilibrium gradient system, the minimum action paths (MAPs) have to go through saddle points. However, for the non-equilibrium working memory system we studied here, the MAPs will deviate from the saddle point (see the MAPs between R and MA state in Fig.~\ref{fig:Fig.3}(C)) due to another force, curl flux~\cite{Li2014PNAS}.
	
	Working memory is desired to be robust to the distractor. The distractor occasionally switches the population dynamics to the distractor-related attractor, followed by incorrect actions. We next analyzed the dynamic properties associated with switching from stimulus-related state (MA) to distractor-related state (MB) under the perturbation of distractor. Both the 90-D transition path (Fig.~\ref{fig:Fig.3}(D)) and the 2-D path after dimension reduction (Fig.~\ref{fig:Fig.3}(E)) reveal that the transition between MA and MB requires passing through R. That is, the former memory needs to be erased first (backing to the resting state) and then a new memory state forms. MA does not suddenly shift to MB without first accessing a state associated with an intermediate cognitive demand. Interestingly, there is a strong positive correlation (r=0.75) between the cortical hierarchy~\cite{Markov2014JCN} and the sequence of deactivation (from high to low firing rate, identified from MAPs (Fig~\ref{fig:Fig.3}(D))) of brain areas during state transition (Fig.~\ref{fig:Fig.3}(F)). This suggests that memory is more robust in association areas compared to early sensory areas, and information flow seems to follow the direction of hierarchical structure. We also found that with the increase of the amplitudes of distractor stimulus, the relative barrier height between MA and MB ($U_{AB}=U_{MA, min}-U_{MB, min}$) increases, i.e., the MA attractor becomes less stable and the MB attractor becomes more stable (Fig.~\ref{fig:Fig.3}(G)). Thus, the state transition from MA to MB is easier, leading to weaker robustness against distractors and poorer behavioral performance.
	
	\subsection{The mechanism of temporal gating of distractors in the delay epoch}
\begin{figure}[htbp]
		\centering
		\includegraphics[width=\linewidth]{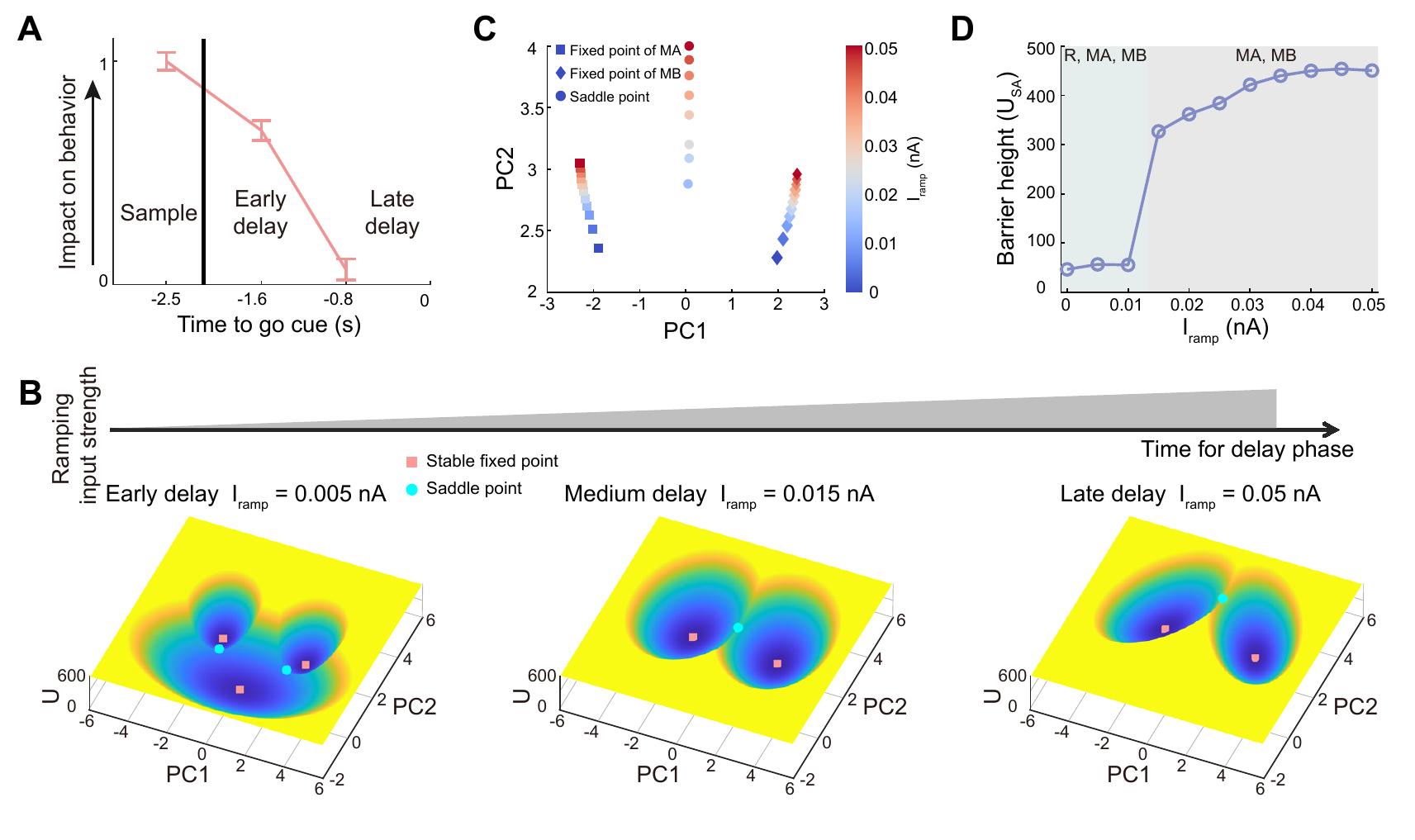}
		\captionsetup{font={stretch=1}, justification=raggedright}
		\newpage
		\caption{\label{fig:Fig.4} (A) Experimentally recorded behavioral impact of distractors delivered at different times of delay epoch in mice. Adapted from~\cite{Finkelstein2021NN}. (B) The time in the delay epoch is modeled as the non-selective ramping external input to all populations in the system~\cite{Inagaki2019Nature,Li2016Nature,Finkelstein2021NN}. Three typical landscapes are presented to illustrate the effect of external ramping input on the attraction basins, corresponding to the distractor delivered at the early, medium, and late time of the delay epoch. (C) The two fixed points for memory states moved further away from each other for stronger ramping input $I_{ramp}$. (D) The barrier height ($U_{SA}$) is defined as the potential difference between saddle point and fixed point of MA increases for increasing ramping input. For weaker ramping input, the distractors can more easily push the system state beyond the saddle point and switch from the target-related attractor to the distractor-related attractor. }
	\end{figure}

	Since there is a temporal separation between sensation and action in the working memory task, to protect the memory against interferences, any additional inputs arriving during the delay epoch need to be effectively gated. As discussed before, the greater strength of distractors is harder to gate. However, recent work reveals that the gating of distractors depends on not only the strength of distractors but also the timing that distractors arrive at~\cite{Finkelstein2021NN}. They found that compared to distractors presented during sample or early delay, distractors presented late in the delay epoch have less impact on behaviors (Fig.~\ref{fig:Fig.4}(A)). That is, distracting stimuli arriving in the delayed epoch become less capable of influencing previous memory as the time to act approaches. As shown below, our landscape approach provides a quantitative explanation for the mechanism of the temporal gating of distractors.
	
	First of all, task-relevant timing is reflected by a continuous, nonspecific ramping input ($I_{ramp}$) to all neural populations in the DWM~\cite{Inagaki2019Nature,Li2016Nature,Finkelstein2021NN}. We quantified three typical attractor landscapes for distractors delivered at the early, medium, and late time of the delay epoch (Fig.~\ref{fig:Fig.4}(B)). For the increase of ramping input (i.e., the later time of delay period), the landscape first shows a phase transition from tristability to bistability and then a progressive separation of attraction basins that encode alternative memories. The results of fixed points and saddle points suggest that the two fixed points move further apart from each other and the distance between fixed point and saddle point increases as the ramping input increases during the delay epoch (Fig.~\ref{fig:Fig.4}(C)). This agrees well with previous results based on the Recurrent Neural Network (RNN)~\cite{Inagaki2019Nature}. Quantitatively, we define barrier height $U_{SA} = U_{S} - U_{MA}$ to quantify the stability of MA, where $U_{S}$ represents the potential at the saddle point formed by MA and its neighboring attractor (R state for tristability and MB state for bistability) and $U_{MA}$ denotes the potential at the minimum of MA attractor. The barrier height is higher for stronger ramping input, indicating that ramping input increases the difficulty of jumping from MA (target-related attractor) to MB (distractor-related attractor) by elevating the corresponding barriers. As a consequence, distractors are more likely to induce a switch between attractors under weaker ramping input compared with stronger input. Therefore, increasing the barrier height of landscape topography results in the filtering of distractors, which provides a quantitative explanation for experimental observations~\cite{Finkelstein2021NN}.

	\subsection{The influence of activation and inhibition strengths on the stability of working memory}
\begin{figure}[htbp]
		\centering
		\includegraphics[width=\linewidth]{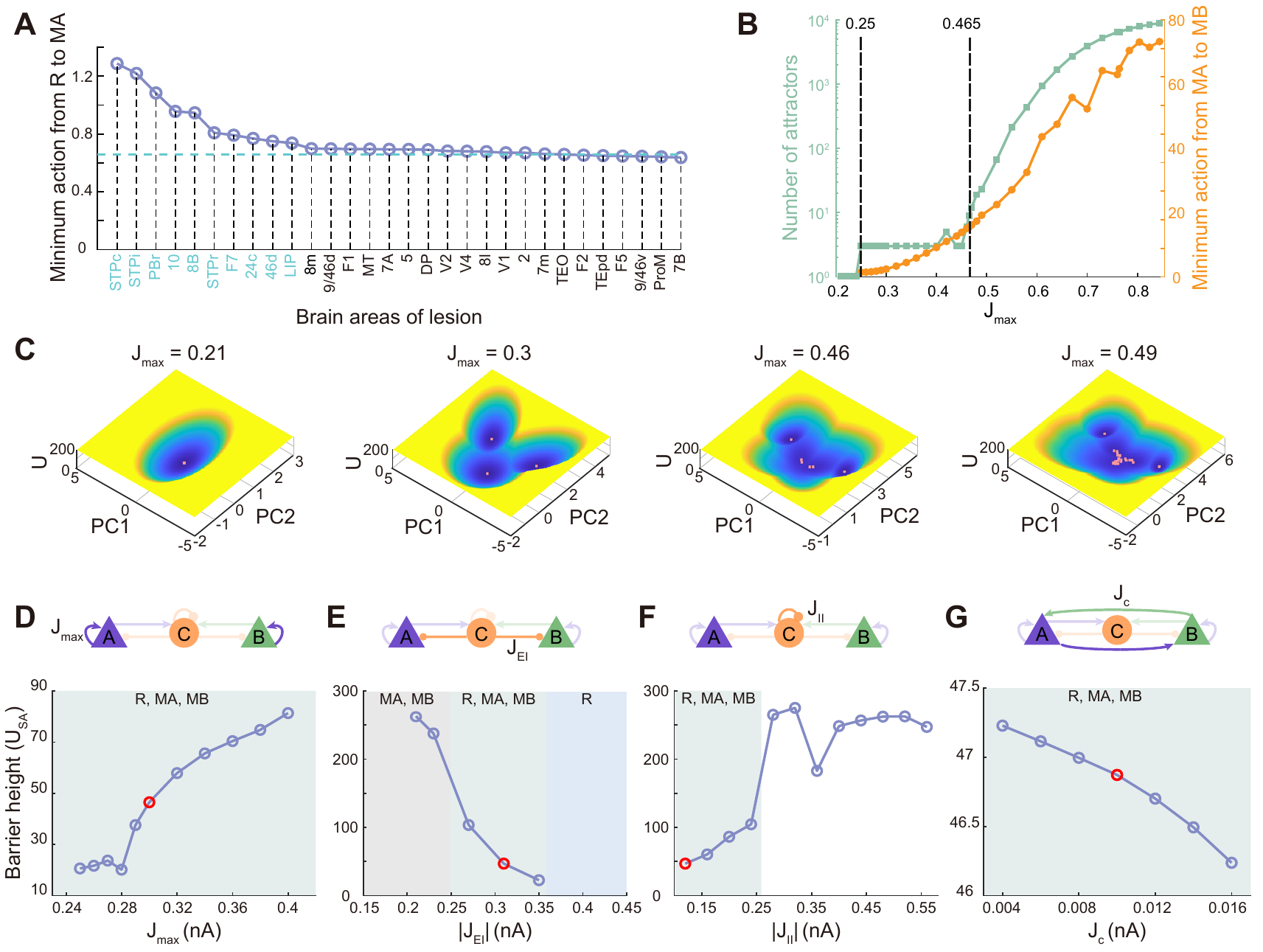}
		\captionsetup{font={stretch=1}, justification=raggedright}
		\caption{\label{fig:Fig.5} (A) Effects of lesions to individual areas on transition action from spontaneous state to the memory state. The blue dashed line represents no brain lesions. The increase of the transition action for any silenced areas suggests that the formation of distributed working memory becomes harder. The top 10 silenced brain areas which have a strong impact on working memory are consistent with the 'bowtie hub' proposed by ~\cite{Markov2013Science} except LIP. (B) Both the number of attractors and transition action from spontaneous state to memory state increase with increasing maximum recurrent strength ($J_{max}$). (C) Four typical landscapes for increasing $J_{max}$. (D-G) The robustness against random fluctuations as a function of the barrier height of MA ($U_{SA}$), quantifying the network structure. The red dots represent default values.}
	\end{figure}

	To explore the roles of different brain areas on working memory function, we can quantify the influence of focal lesions on individual areas on the robustness of distributed working memory patterns by the minimum action from R to MA under no external stimulation. The silence of the brain area is conducted by cutting off its communication with other areas. The increase (decrease) of action after silencing an individual area suggests that the formation of distributed working memory becomes harder (easier) after silencing the corresponding brain area, which demonstrates that this area plays key a role in maintaining the working memory state. The blue line in Fig.~\ref{fig:Fig.5}(A) represents no brain lesions. We picked the top 10 key brain areas identified from the above procedure which display a strong impact on maintaining working memory function, and found that these 10 areas are consistent with the 'bowtie hub' proposed by ~\cite{Markov2013Science}.
	
	Across all analyses performed above, we assumed the maximum area-specific recurrent synaptic strength $J_{max}=0.3$, below the critical value needed for tristability in isolation (0.465). We explored the change of the number of attractors and minimum action from MA to MB as a function of $J_{max}$ in Fig.~\ref{fig:Fig.5}(B) ($J_{min}$ is fixed as 0.21). The number of attractors is determined by initializing the DWM by 10000 initial conditions sampled from standard uniform distribution and counting the number of unique fixed points using a tolerance of 10\%. There exists a lower limit of $J_{max}$ (0.25 in our simulation) for the emergence of memory-related attractors (compare the first two subfigures in Fig.~\ref{fig:Fig.5}(C)), highlighting the importance of the gradient of synaptic excitation in the neocortex. When $J_{max}$ exceeds the threshold, larger $J_{max}$ indicates that more brain areas high in the hierarchy have strong local reverberation to generate self-sustained persistent activity and be able to display multistability even isolated from the network. Therefore, it is reasonable that the large-scale circuit we studied here displays a large number of attractors for larger $J_{max}$ (green line in Fig.~\ref{fig:Fig.5}(B)). Some typical landscapes with multistable attractors are presented in Fig.~\ref{fig:Fig.5}(C). These novel attractors potentially subserve internal processes and need to be validated by experiments in the future. Furthermore, the minimum action from MA to MB tends to increase with respect to $J_{max}$, demonstrating the more difficult state transition and thus stronger robustness of the memory state.
	
	Human cognitive processes including working memory are influenced not only by external task demands but also by inherent fluctuations. During the delay period, if there is no external distractor, the robustness of working memory is influenced by the non-specific noise. The robustness against noise relies on both the magnitude of noise and the network structure. It is anticipated that a larger magnitude of noise leads to weaker robustness. In a previous study on the classic local circuit model of working memory, it is found that both increasing self-excitation and mutual inhibition are beneficial to the robustness of the working memory system to the random fluctuations~\cite{Yan2020PlosCB}. However, for the distributed working memory model engaging multiple interacting brain areas we studied here, the role of network structure on robustness is more complex. In Fig.~\ref{fig:Fig.5}(D-G), we show the dependence of the barrier height of MA ($U_{SA}$) on the network structure (the red dots represent the case for default modeling parameters), more specifically, on the maximum area-specific recurrent synaptic strength ($J_{max}$), the inhibition strength from inhibitory population to excitatory populations ($J_{EI}$), the self-inhibition strength of inhibitory population ($J_{II}$) and the cross-coupling strength between excitatory populations ($J_{C}$). The MA state with higher barrier height (larger $U_{SA}$) is more robust to the noise.
	
	For $J_{max}$, we focus on the situation where the landscape is tristable, and we found that as $J_{max}$ increases, the attractors representing memory states become deeper while the attractor representing resting state becomes shallower, leading to harder noise-driven switch from memory attractor to the resting one (Fig.~\ref{fig:Fig.5}(D)). This is in line with the results from minimum action (orange line in Fig.~\ref{fig:Fig.5}(B)) indicating larger transition action for larger $J_{max}$. This is also consistent with the previous study based on local circuit~\cite{Yan2020PlosCB}. The influence of $J_{EI}$ is more complex (Fig.~\ref{fig:Fig.5}(E)). For small $J_{EI}$, the landscape exhibits two symmetric attractors encoding memory states. The additional increase of $J_{EI}$ induces the emergence of the attractor representing the resting state. However, the two symmetric memory states disappear simultaneously for further increase of $J_{EI}$, so the system loses its ability to perform working memory. The decaying $U_{SA}$ as a function of $J_{EI}$ suggests that weaker inhibition strength to excitatory populations is requisite for the emergence and strong robustness of memory states. Consistently, the self-inhibition of inhibitory populations is required to be stronger to reduce the inhibition to excitatory populations, thus enhancing the robustness of memory states (Fig.~\ref{fig:Fig.5}(F)). Furthermore,  smaller mutual excitation between two competitors (population A and B) is beneficial to the winner-take-all competition between the two competitors, explaining the declining $U_{SA}$ for larger $J_{C}$ (Fig.~\ref{fig:Fig.5}(G)).
	
	\section{Discussion}
	Traditionally, the investigation of cognitive functions is limited to operations on local brain circuits. However, with the advances of both experimental recording techniques~\cite{Jun2017nature} and mathematical large-scale cortical models~\cite{Schmidt2018PlosCB,Chaudhuri2015Neuron}, uncovering the underlying mechanisms of distributed cognition becomes increasingly feasible. Recently, some mathematical modeling efforts have been devoted to exploring the distributed working memory in macaque~\cite{Mejias2022Elife,Froudist2021Neuron}. The simulation results agree well with experimental observations of multiple cortical areas participating in working memory tasks~\cite{Leavitt2017,Christophel2017TCS,Sreenivasan2019NRN}. Some novel phenomena, such as the separation between the areas displaying sustained activity and those
	that did not and counterstream inhibition, emerge in large-scale network~\cite{Mejias2022Elife,Froudist2021Neuron}. The large-scale model engaging multiple brain regions also facilitates the investigation of the contributions of different brain regions to distributed functions.
	
	Working memory is associated with the stimulus-selective persistent activity after the withdrawal of stimulus. The storage of memory requires the robustness of persistent activity to inherent noise in brains and external distractors. The concept of attractor dynamics was proposed by previous theoretical works to describe the robustness of working memory~\cite{Murray2017JN,Seeholzer2019PlosCB,Inagaki2019Nature}. In the attractor landscape or energy landscape framework, each possible stable state corresponds to a basin of attraction, and the stability of each stable state can be quantified by the depth of the basins. Although some attempts have been made to explicitly quantify the attractor landscape~\cite{Yan2020PlosCB,Ye2021Frontiers}, such landscapes were mostly based on local brain circuits, rather than large-scale networks recruiting multiple brain areas. Different from exploring the robustness of working memory through tracking the neural activity trajectories in a limited time scale~\cite{Mejias2022Elife}, which mostly reflects the local properties, the landscape approach can reveal the global nature of working memory. In addition, for the large-scale network, how to visualize the landscape remains a challenge.
	
	In this work, based on the ordinary differential equations of anatomically constrained distributed working memory model developed by~\cite{Mejias2022Elife}, we quantified the attractor landscape with moment equation approximations~\cite{Li2014PNAS,li2013PLOS,Kang2021AS} and visualized it with dimension reduction approach~\cite{Kang2021AS}. The landscape provides a global picture of the stochastic dynamics of distributed working memory. In the absence of external stimulation, three attractors were identified in the subspace spanning by PC1 and PC2, i.e., spontaneous state and two memory states encoded by two selective excitatory populations, respectively. The working memory function is governed by the change of landscape topography and switch of system state according to the task paradigm, which explains the neural activities trajectories from multiple brain regions. The noise-driven state transition from stimulus-related attractor to distractor-related attractor during the delay epoch accounts for the behavioral error.
	
	In this investigation, two kinds of noises that may induce state transition are considered: non-selective random fluctuations and distractor stimuli. By quantifying the landscape topography through barrier heights, we found that larger strength of distractor stimuli destabilizes stimulus-related attractors, leading to weaker robustness to distractors. For non-selective fluctuations, we explored the dependence of the robustness of the working memory state on the network structure, more specifically, the excitation and inhibition strengths in local circuits. We found that both stronger strength of self-excitation and self-inhibition are beneficial to the robustness of the working memory state while stronger strength of inhibition to excitatory populations and cross-coupling between excitatory populations have the opposite effect. These results provide new insights into the design of network structure to achieve the better behavioral performance of cognitive functions.
	
	Furthermore, the most probable biological path identified by the minimum action path approach reveals that the spontaneous state serves as an intermediate state during the switch between the two memory states, which may increase the plasticity of state transition. The memory stored in the cortical area with higher hierarchy is much more difficult to lose, and thus more stable. The increasing barrier height of stimulus-related attractors with time-evolving during the delay epoch also explains the mechanism of the temporal gating of distractors. Interestingly, many novel attractors emerge during the increase of self-excitation, which may serve as various forms of internal representations. The functional role of these attractors warrants further explorations from both experimental and theoretical efforts.
	
	As a complex system, the brain exhibits distinct principles of organization operating at different scales and the large-scale activity of the brain is more than the trivial sum of its components~\cite{Breakspear2005PTRSB}. Thus, biophysical models of large-scale neuronal activity and their dynamics are crucial for the understanding of cognitive function~\cite{Breakspear2017NN}. Our energy landscape with dimension reduction and transition path approach provides a new paradigm to study the dynamics of large-scale neural networks. Here, we explored the underlying transition dynamics of distributed working memory, which is general to many other cognitive functions such as distributed decision-making. In the future, it is expected that the landscape approach can be applied to study the underlying mechanism of other cognition functions and the brain disorders such as schizophrenia, which are often related to significant cognitive impairment including working memory. Besides, combining the large-scale brain model with multimodal neuroimaging data is also essential for accurately modeling and explaining the mechanisms of
	human brain function in health and disease~\cite{Deco2019PNAS,Kringelbach2020PNAS}.
	
	\section{Methods}
	We provide an overview of the workflow for the large-scale model, the energy landscape construction, and transition path quantification based on the distributed working memory network.
	
	\subsection{The anatomically constrained large-scale model of distributed working memory in macaque}
	The mathematical model of distributed working memory we explored here was first proposed and carefully described in~\cite{Mejias2022Elife}. Here we briefly summarize the model.	Fig.~\ref{fig:Fig.1}(A) presents the schematic diagram of the large-scale model of the macaque cortex, which engages cortical areas across all four neurocortical lobes, including V1, V2, MT, LIP, V4, 7A, 7m, 8m, 8l, 5, TEO, DP, 2, F1, 7B, TEpd, 10, F5, 46d, PBr, 24c, F2, ProM, STPc, STPi, STPr, F7, 8B, 9/46v, 9/46d. Each cortical area is modeled as the of Wong-Wang model with three fully connected populations~\cite{Wong2006JNeurosci}. The temporal evolution of the two selective excitatory populations (labeled as A and B) and one inhibitory population (labeled as C) is described by the following equations~\cite{Mejias2022Elife}:
	\begin{gather}
		\frac{d S_{A}}{d t}=-\frac{S_{A}}{\tau_{N}}+\gamma_{E} \left(1-S_{A}\right) r_{A}+\zeta_{A}(t), \\
		\frac{d S_{B}}{d t}=-\frac{S_{B}}{\tau_{N}}+\gamma_{E} \left(1-S_{B}\right) r_{B}+\zeta_{B}(t), \\
		\frac{d S_{C}}{d t}=-\frac{S_{C}}{\tau_{G}}+\gamma_{I} r_{C}+\zeta_{C}(t), \\
		I_{A}=J_{S} S_{A}+J_{C} S_{B}+J_{E I} S_{C}+I_{0 A}+I_{n e t}^{A}, \\
		I_{B}=J_{C} S_{A}+J_{S} S_{B}+J_{E I} S_{C}+I_{0 B}+I_{n e t}^{B}, \\
		I_{C}=J_{I E} S_{A}+J_{I E} S_{B}+J_{I I} S_{C}+I_{0 C}+I_{n e t}^{C}, \\
		r_{A, B}(I_{A, B})=\frac{a I_{A, B}-b}{1-\exp [-d(a I_{A, B}-b)]}, \\
		r_{C}(I_{C})=\left[\frac{1}{g_{I}}\left(c_{1} I_{C}-c_{0}\right)+r_{0}\right]_{+}.
	\end{gather}
	Here, $S_A$ and $S_B$ are interpreted as the fraction of open NMDA-mediated synaptic channels in populations A and B, and $S_C$ is the fraction of open GABAergic-mediated synaptic channel in population C. $I_{i}$ with i = A, B, C is the synaptic current input to the population 'i', which is composed of local inputs from local circuits, background inputs ($I_{0i}$) and long-range inputs from other areas in the network ($I_{net}^{i}$).  $r_i$ is the population-averaged firing rate. The notation $[\cdot]_{+}$ in $r_C$ denotes rectification. $\zeta_i$ is the Gaussian white noise, introducing stochasticity into the system.
	
	\begin{table}[tbp]\centering
		\caption{Parameters used for modeling distributed working memory network.}
		\label{tab:table1}%
		\begin{tabular}{p{2cm}|p{4.5cm}|p{10cm}}
			\textbf{Parameter} & \textbf{Default Value} & \textbf{Description}\\
			\hline
			$\tau_{N}$ & $60$ ms &  Time constant of NMDA receptor \\
			$\tau_{G}$ & $5$ ms & Time constant of GABAergic receptor \\
			$\gamma_{E}$ & $1.282$ & $-$ \\
			$\gamma_{I}$ & $2$ & $-$ \\
			$J_{min}$ & $0.21$ & Self-excitation strength of excitatory populations for brain area with lowest hierarchy \\
			$J_{max}$ & $0.3$ & Self-excitation strength of excitatory populations for brain area with highest hierarchy \\
			$h$ & $-$  & Hierarchy of 30 brain areas \\
			$J_{S}$ & $J_{min} + (J_{max}-J_{min})h$ & Self-excitation strength of excitatory populations in 30 brain areas\\
			$J_{C}$ & $0.0107$ nA & Cross-coupling strength
			between excitatory populations \\
			$J_{EI}$ & $-0.31$ nA & Coupling strength from inhibitory population to excitatory ones \\
			$J_{IE}$ & Area-specific & Coupling strength from the excitatory populations to the inhibitory one \\
			$J_{II}$ & $-0.12$ nA & Self-inhibition
			strength of the inhibitory population \\
			$I_{I0A}$ & $0.3294$ nA & Background inputs to population A \\
			$I_{I0B}$ & $0.3294$ nA & Background inputs to population B \\
			$I_{I0C}$ & $0.26$ nA & Background inputs to population C \\
			$a$, $b$, $d$ & 135 Hz/nA, 54 Hz, 0.308 s  & Parameters for the transfer function of excitatory populations\\
			$g_{I}$, $c_{1}$, $c_{0}$, $r_{0}$ & 4, 615 Hz/nA, 177 Hz, 55 Hz  & Parameters for the transfer function of inhibitory populations\\
			$G$ & $0.48$  & Global coupling strength \\	 	
		\end{tabular}
	\end{table}
	
	To introduce area-to-area heterogeneity, it is assumed that the local recurrent strength from the early sensory area to the higher association area increases in a gradient way. This is inspired by the anatomical study which identified a gradient of dendritic spine density~\cite{Elston2007}. In addition, the cortical areas are distributed along an anatomical hierarchy. The anatomical fraction of supragranular layer neuron data is used to build the anatomical hierarchy. The low (high) hierarchy of the source area relative to the target corresponds to the high (low) SLN values of source-to-target projection~\cite{Felleman1991CerebCorte,Markov2014JCN}. Area $i$ is assigned a normalized hierarchical value $f_{i}$ estimated by logistic regression using FLN~\cite{Markov2014JCN}. More specifically, the local recurrent strength of area $i$ is given by
	\begin{equation}
		J_{S}(i)=J_{\min }+\left(J_{\max }-J_{\min }\right) h_{i},
	\end{equation}
	where $h_{i}$ is the normalized value between zero and one of the dendritic spine count values with age-related corrections observed in anatomical studies. For the missing value of spine count, the anatomical value $f$ is used as a proxy due to the high correlation between spine count data and anatomical hierarchy. Therefore, the value of $J_{S}$ varies from $J_{min}$ to $J_{max}$. Furthermore, to ensure that the spontaneous activity is the same for all areas, $J_{IE}$ also scales with $J_{S}$:
	\begin{equation}
		J_{I E}=\frac{1}{2 J_{E I} \zeta}\left(J_{0}-J_{S}-J_{C}\right),
	\end{equation}
	where $\zeta=\frac{\tau_{G} \gamma_{I} c_{1}}{g_{I}-J_{I I} \tau_{G} \gamma_{I} c_{1}}$ and $J_{0}=0.2112$ nA. The minimum value of $J_{min}$ is 0.205 nA to ensure $J_{IE}$ is non-negative. In this work, $J_{min}$ is set to be 0.21. $J_{max}$ is an important parameter, which determines the type of working memory model. Fig.~\ref{fig:Fig.1}(B) shows the bifurcation diagram of the isolated area with respect to local coupling $J_{S}$. When $J_{max}$ is below the bifurcation point, all areas are monostable in isolation. In this case, it is the inter-areal projections that contribute to the emergence of sustained activity displayed by the model. On the other hand, having $J_{max}$ above the bifurcation point implies that some higher cognitive areas are intrinsically multistable when isolated, compatible with classical working memory theories.
	
	For inter-areal projections, the quantitative anatomical connectivity data in macaque is available by track-tracing studies~\cite{Markov2013Science,Markov2014CerebCorte,Markov2014JCN}. The fraction of labeled neurons (FLN) is defined as a proxy of the connection strength between cortical areas. The FLN, as we have discussed before, defines the structural hierarchy of the network. Both FLN and SLN are weighted and directed matrices. The connectivity data can be downloaded from \url{https://core-nets.org}.
	
	Since the inhibitory projections tend to be local, it is assumed that only excitatory populations can generate inter-areal projections. To facilitate the propagation of activity along the hierarchy, it is also assumed that the long-distance outgoing connectivity targets more strongly excitatory populations in a selective way for feedforward pathways and inhibitory populations for feedback pathways (Fig.~\ref{fig:Fig.1}(A)). Then the long-range input to the three populations of area $x$ originated from all the other cortical areas $y$ is given by
	\begin{gather}
		I_{A, n e t}^{x}=G \frac{J_{S}(x)}{max(J_{S})} \sum_{y} W^{x y} S L N^{x y} S_{A}^{y}, \\
		I_{B, n e t}^{x}=G \frac{J_{S}(x)}{max(J_{S})} \sum_{y} W^{x y} S L N^{x y} S_{B}^{y}, \\
		I_{C, n e t}^{x}=\frac{G}{Z} \frac{J_{IE}(x)}{max(J_{IE})} \sum_{y} W^{x y}\left(1-S L N^{x y}\right)\left(S_{A}^{y}+S_{B}^{y}\right).
	\end{gather}
	Here, $G$ is the global coupling strength, taken as 0.48 unless specified otherwise. $Z$ is a factor balancing long-range excitatory and inhibitory projections, take as $Z=\frac{2 c_{1} \tau_{G} \gamma_{I} J_{E I}}{c_{1} \tau_{G} \gamma_{I} J_{I I}-g_{I}}$ to guarantee the net effect of population A and B with the same activity level on other areas is zero. The sum of $y$ is for all 30 cortical areas of the network. For population A in a specific area, it is influenced by the A-selective populations of other areas directly and B-selective populations of other areas indirectly via local population C.
	
	Due to the board range of FLN values, FLN is firstly rescaled to a suitable range for the firing rate model by $W_{xy}=k_{1}(FLN^{xy})^{k_{2}}$ where $k_{1}=1.2$ and $k_{2}=0.3$. Then $W$ is furthered normalized so that $ \sum_{y} W^{x y}=1$. Finally, the gradient is also introduced into the long-range projection strengths by multiplying the gradient of $J_S$ for feedforward projections and $J_{IE}$ for feedback projections. The SLN is taken as the proxy of the feedforward and feedback characteristics of projections. That is, $SLN=1$ stands for pure feedforward network and $SLN=0$ represents pure feedback network. So the linear dependence on SLN for projections to excitatory populations and (1-SLN) for projections to inhibitory is also assumed. Inspired by the evidence that frontal networks have primarily strong excitatory loops~\cite{Markowitz2015PNAS}, the SLN-driven modulation of feedback projections from frontal areas to 8l and 8m is modified to be no larger than 0.4. The default values of all parameters and their descriptions are shown in Table 1.

	\subsection{Quantification of the energy landscape with moment equations}
	
	For the high dimension and nonlinear dynamical system of distributed working memory model described above, the numerical solution can be obtained by the Euler's Method with an integration time step of $0.01 ms$. Furthermore, the time evolution of stochastic dynamics can be expressed in vector form as
	\begin{equation}
		\frac{d \bm{x}(t)}{dt} = \bm{F}(\bm{x}(t)) + \bm{\zeta}(t),
		\label{eq:LE}
	\end{equation}
	where $\bm{x}(t)=[S_{A, V1}(t), S_{B, V1}(t), S_{C, V1}(t), ...,S_{A, 9/46d}(t), S_{B, 9/46d}(t), S_{C, 9/46d}(t)]$, representing the time-varying 90-D system variables. ${\bm F}({\bm x})$ denotes the driving force of the system. $\bm{\zeta}$ is a 90-D vector representing independent Gaussian white noise, that is,
	\begin{gather}
		<{\bm{\zeta}} (t)> = 0, \\
		<{\bm{\zeta}} (t) {\bm{\zeta}}(t')> = 2 d \cdot {\bm I} \cdot \delta (t-t'),
	\end{gather}
	where $\bm{I}$ is the identity matrix and $\delta (t)$ is Dirac delta function. $d$ is the constant diffusion coefficient, characterizing the level of noise.
	
	The corresponding diffusion equation (Fokker–Planck equation) of Eq.~\ref{eq:LE} describing the time evolution of the system probability distribution takes the form as
	\begin{equation}
		\frac{\partial p(\bm{x}, t)}{\partial t}=-\sum_{i} \frac{\partial}{\partial x_{i}}\left[F_{i}(\bm{x}, t) p(\bm{x}, t)\right]+d \sum_{i} \sum_{j} \frac{\partial^{2}}{\partial x_{i} \partial x_{j}} p(\bm{x}, t),
		\label{eq:FPE}
	\end{equation}
	where $p(\bm{x}, t)$ is the probability density function of system state at time $t$. However, due to the high dimension (90-D) and nonlinearity, the partial differential equation of Eq.~\ref{eq:FPE} is intractable. So we assumed that the time evolution of the density function satisfies the Gaussian distribution. When the diffusion coefficient $d$ is small, the moment equations describing the time evolution of mean $\bar{\bm{x}}(t)$ and variance matrix $\bm{\Sigma}(t)$ of Gaussian distribution are as follows~\cite{hu1994stochastic,Van2007stochastic}:
	\begin{gather}
		\dot{\bar{\bm{x}}}(t)=\bm{F}[\overline{\bm{x}}({t})], \label{eq:mean}\\
		\dot{\bm{\Sigma}}(t)=\bm{\Sigma}(t)\bm{A}^{T}(t)+\bm{A}(t)\bm{\Sigma}(t)+2{d} \cdot \bm{I}.
		\label{eq:variance}
	\end{gather}
	Here, $\bm{A}(t)$ is the Jacobian matrix of $\bm{F}(x)$, i.e., $A_{i j}(t)=\frac{\partial F_{i}(\overline{\bm{x}}(t)}{\partial x_{j}}$, and $\bm{A}^T(t)$ is the transpose of $\bm{A}(t)$.
	
	In this work, we focus on the behaviors of the system at the steady state. The probability density distribution of the system at the steady state can be expressed as:
	\begin{equation}
		p_{ss}(\bm{x})=\frac{1}{(2 \pi)^{\frac{N}{2}}|\bm{\Sigma}|^{1 / 2}} \exp \left\{-\frac{1}{2}(\bm{x}-\bm{\mu})^{T} \bm{\Sigma}^{-1}(\bm{x}-\bm{\mu})\right\},
		\label{eq:gaussian}
	\end{equation}
	where $\bm{\mu}$ and $\bm{\Sigma}$ are the solutions of Eq.~\ref{eq:mean} and Eq.~\ref{eq:variance} when $t\rightarrow +\infty$, respectively. Dependent on the initial conditions of the system variable (the conductance of receptors in this work), Eq.~\ref{eq:mean} and Eq.~\ref{eq:variance} may have different solutions at the steady state, which means that the system may be multistable. In this condition, the final probability density function is determined by the weighted sum of multiple Gaussian distributions, that is,
	\begin{equation}
		p_{ss}(\bm{x})=\sum_{j=1}^{M} w_{j} p^{j}_{ss}(\bm{x}).
		\label{eq:pss}
	\end{equation}
	Here, $M$ is the number of stable states of the system, $p^j_{ss}(\cdot)$ represents the density function of $j$th stable state taken the form of Eq.~\ref{eq:gaussian} and $w_{j}$ denotes the corresponding weight. The weight is estimated by the statistic of frequency of each stable state under abundant initial conditions (10000 initial conditions in this work). The system variables are randomly initialized by sampling from the uniform distribution in the range from 0 to 1.
	
	Finally, the potential landscape at the steady state is defined as $U(\bm{x})=-\ln P_{ss}(\bm{x})$~\cite{Li2014PNAS,li2013PLOS,Kang2021AS}, where $P_{ss}$ represents the normalized probability distribution at the steady state, and $U$ is the dimensionless potential. However, for the high-dimensional system, the high-dimensional potential landscape is hard to visualize and interpret. We used previously developed dimension reduction approach to display the landscape in reduced dimensions ~\cite{Kang2021AS}.
	
	\subsection{The dimension reduction of high-dimensional potential landscape}
	Inspired by the principle of Principal Component Analysis (PCA), we can reduce the dimensionality of the potential landscape but at the same time minimize information loss by projecting the system variables $\bm{X}$ into the new subspace spanning by the first few principal components~\cite{Kang2021AS}. More specifically, for the probability density function of the multistable system of Eq.~\ref{eq:pss}, the mean and variance of $\bm{X}$ are
	\begin{gather}
		\bm{\mu} = \sum_{j=1}^{M} w^{j} \bm{\mu}^{j}, \\
		\bm{\Sigma} = \sum_{j=1}^{M} w^{j} (\bm{\Sigma}^{j} + \bm{\mu}^{j} (\bm{\mu}^{j})^{T}) - \bm{\mu} \bm{\mu}^{T}.
	\end{gather}
	Since the covariance matrix $\bm{\Sigma}$ is the positive semi-definite matrix, we can perform eigenvalue decomposition on $\bm{\Sigma}$. The eigenvectors corresponding to the first $C$ largest eigenvalues are denoted by $\bm{V} = (\bm{v}_1, ..., \bm{v}_C)$ with $\Vert \bm{v}_i \Vert=1$  and $ \bm{v}_i \bm{v}_j^{T}=0$, $ \forall i\neq j$. The projected system variable $\bm{Z}$ on the subspace spanning by $\bm{V}$ is denoted by $\bm{Z}=\bm{V}^{T}\bm{X}^j=(Z_{1}, ..., Z_{C})$. The mean and covariance matrix of $j$th stable state after projection are $\bm{\mu}_{z}^{j}=\bm{V}^{T}\bm{\mu}_{j}$ and  $\bm{\Sigma}_{z}^{j}=\bm{V}^{T}\bm{\Sigma}^{j}\bm{V}$, respectively. Correspondingly, the multi-dimensional normal distribution of $j$th stable state becomes
	\begin{equation}
		p_{z}^{j}\left(\mathbf{z}\right) =\frac{1}{(2 \pi)^{\frac{C}{2}}\left|\bm{\Sigma}_{z}^{j}\right|^{\frac{1}{2}}} \exp \left\{-\frac{1}{2}\left(\mathbf{z}-\bm{\mu}_{z}^{j}\right)^{T}\left(\bm{\Sigma}_{z}^{j}\right)^{-1}\left(\mathbf{z}-\bm{\mu}_{z}^{j}\right)\right\}.
	\end{equation}
	The final probability density function after dimension reduction is $p_z(\bm{z})=\sum_{j=1}^{M} w_{j} p^{j}_{z}(\bm{z})$, and the potential landscape is $U_z(\bm{z})=-ln(P_z(\bm{z}))$. In this work, we choose $C=2$, that is, the high-dimensional potential landscape is projected to the 2D subspace spanning by the first two components $PC1$ and $PC2$.
	
	\subsection{Identification of minimum action paths (MAPs)}
	
	For a multistable system, it is of great importance to identify the transition paths between different attractors, which provide the information on switching order for different components in the state transition process. Firstly, we define the path between $i$th attractor $\bm{x}^i$ at time 0 and $j$th attractor $\bm{x}^j$ at time $T$ as $\bm{x}^{i j}(t)=\left(x_{1}^{i j}(t), x_{2}^{i j}(t), \cdots, x_{n}^{i j}(t)\right)^{T}$, $t \in[0, T]$. Then the path $\bm{x}^{i j}(t)$ satisfies
	the following boundary conditions:
	\begin{equation}
		\left\{\begin{array}{c}
			\left(x_{1}^{i j}(0), x_{2}^{i j}(0), \cdots, x_{N}^{i j}(0)\right)^{T}=\bm{x}^{i}, \\
			\left(x_{1}^{i j}(T), x_{2}^{i j}(T), \cdots, x_{N}^{i j}(T)\right)^{T}=\bm{x}^{j}.
		\end{array}\right.
	\end{equation}
	
	Let $L^{ij}$ be the distance between the driving force $\bm{F}$ and the velocity along the transition path: $L^{i j}\left(t, \bm{x}^{i j}(t), \frac{d \bm{x}^{i j}(t)}{d t}\right)=\left\|\left(\frac{d x_{1}^{i j}(t)}{d t}, \frac{d x_{2}^{i j}(t)}{d t}, \cdots, \frac{d x_{N}^{i j}(t)}{d t}\right)^{T}-\bm{F}\left(\bm{x}^{i j}(t)\right)\right\|_{2}$. The transition action $S(\bm{x}^{ij})$ is defined as the integral of the Lagrangian $L^{ij}$ between time 0 and $T$. Based on Wentzell-Freidlin theory~\cite{Freidlin2012}, the estimation of the probability distribution of the solution $\bm{x}(t)$ over time interval $[0, T]$ at a give $\delta$ is
	\begin{equation}
		P\left\{\rho\left(\bm{x}(t), \bm{x}^{i j}(t)\right)<\delta\right\} \approx \exp \left(-\frac{S_{i j}\left(\bm{x}^{i j}\right)}{\varepsilon}\right).
	\end{equation}
	So the most probable transition path starting from $\bm{x}^i$ for $t=0$ and ending at $\bm{x}^j$ for $t=T$ can be obtained by solving the following optimal problem~\cite{E2004,Zhou2008JCP,Heymann2008}:
	\begin{equation}
		\min _{\bm{x}^{i j}} S\left(\bm{x}^{i j}\right)=\frac{1}{2} \min _{\bm{x}^{i j}} \int_{0}^{T} L^{i j}\left(t, \bm{x}^{i j}(t), \frac{d \bm{x}^{i j}(t)}{d t}\right)	d t. \label{eq:path}
	\end{equation}
	This optimal path is called the minimum action path (MAP), which is also identified as the biological path between attractors. In this work, we calculated the MAPs numerically, using the minimum action methods~\cite{Zhou2008JCP,Li2018npj} (See Supplementary Material for details).
	
	\section*{Acknowledgement}
	C.L. is supported by the National Key R\&D Program of China (2019YFA0709502) and the National Natural Science Foundation of China (12171102).
	
	\newpage
	\bibliography{DWM}
	
	\newpage
	
\end{document}